\documentclass[trackchanges,twocolumn,longauthor]{aastex701}
\shorttitle{CROCS Data Release I}
\shortauthors{The CROCS collaboration}
\received{April 1, 2024}
\revised{April 1, 2025}
\accepted{April 1, 2026}
\submitjournal{Monthly Notices of the Royal Astrological Society}
\graphicspath{{./}{figures/}}

\usepackage{caption}
\usepackage{subcaption}
\usepackage{physics}
\usepackage{derivative}
\usepackage{tcolorbox}
\usepackage{censor}

\begin{document}

\title{CROCS\footnote{Cosmologists Researching On Cosmological Scales} Data Release I: Constraints on the Hubble Constant}


\author[orcid=0000-0003-1175-8004,gname=Luke,sname=Weisenbach]{Luke Weisenbach}
\altaffiliation{Corresponding author: available upon lunar alignment}
\altaffiliation{Middle of Nowhere, Indiana}
\altaffiliation{Office 2.07} 
\affiliation{LBC, UK}
\email[show]{weisluke@alum.mit.edu} 

\author[orcid=0009-0001-3422-3048,gname=Firstname, sname=Lastname]{Sophie L. Newman} 
\affiliation{Department of Observational Guesswork, A Generic University, UK} 
\email[show]{sophie.newman@port.ac.uk}

\author[orcid=0000-0000-0000-0000,gname=Firstname, sname=Lastname]{Kieran Graham} 
\altaffiliation{Office 2.07}
\affiliation{LBC, UK} 
\email[show]{kieran.graham@port.ac.uk}

\author[orcid=0009-0008-9176-7974,gname=Firstname, sname=Lastname]{Sai S. Dhavala} 
\altaffiliation{These authors consented to having their names added to the author list for the purpose of enbiggening the paper}
\affiliation{Land of Daydreaming, Some Other Planet, Orion Arm, MW} 
\email[show]{sai.srinivas@port.ac.uk}

\author[orcid=0000-0003-4175-571X, gname=Benjamin, sname=Floyd]{Benjamin Floyd}
\altaffiliation{The Place Over There, You Know, That One}
\affiliation{CROCS Collaboration Headquarters, Swamp Analytics Division, Florida, USA}
\affiliation{Institute of Cosmology and Gravitation, University of Portsmouth, UK}
\affiliation{Faculty of Physics and Astronomy, University of Missouri--Kansas City, USA}
\email[show]{benjamin.floyd@port.ac.uk}

\author[orcid=0000-0000-0000-0000,gname=Firstname, sname=Lastname]{Neel Shah}
\altaffiliation{These authors consented to having their names added to the author list for the purpose of enbiggening the paper} 
\affiliation{Institute for Advanced Unit Conversion Studies, Metric College, UK} 
\email[show]{astronomer@astronomy.space} 

\author[orcid=0000-0000-0000-0000]{Gemini 3 Flash}
\altaffiliation{Somewhere in the ``cloud''}
\affiliation{Google}
\email[show]{gemini.google.com}
\collaboration{all}{The CROCS Collaboration}

\begin{abstract}

Recent cosmological surveys and datasets have highlighted a variety of tensions\footnote{We note that we do not consider the so-called `sigma 8 ($\sigma_8$) tension' to be an actual tension.} to the concordance model of our universe, $\Lambda$CDM. Of particular interest is the Hubble tension, the $5.5\sigma$ discrepancy between measurements of the Hubble constant $H_0$ using high redshift CMB data from Planck ($67.27\pm0.60$km$\text{s}^{-1}\text{Mpc}^{-1}$) and low redshift supernovae from SH0ES ($73.2\pm1.3$km$\text{s}^{-1}\text{Mpc}^{-1}$). To avoid stepping on any toes, we have initiated the CROCS collaboration to resolve this tension, gathering experts from across many fields of cosmology, astrophysics, astronomy, machine learning, data science, philosophy, and astrology. In this paper, we present findings from CROCS Data Release 1, corresponding to the first $\sim3$ days and 27 minutes (rest frame) of observation. We perform a robust statistical analysis, showing that Planck and SH0ES both suffer from imperial biasing systematics (IBS) at $5\sigma$ significance. Accounting for these errors by converting to metric units reconciles the high and low redshift data, with $H_0 = 69.00\pm0.420$km$\text{s}^{-1}\text{Mpc}^{-1}$. We thus report that our results are sufficient to end the Hubble tension for good.  

\end{abstract}



\section{Introduction} 

The Hubble tension is the $5.5\sigma$ discrepancy between two independently measured values of the Hubble constant $H_0$. The Planck collaboration, analysing CMB temperature anisotropies in the early universe, reports $H_0 = 67.27\pm0.60$\,km\,s$^{-1}$\,Mpc$^{-1}$ \citep{PlanckCollaboration:2020}, while the SH0ES collaboration, using Cepheid-calibrated Type Ia supernovae in the late universe, find $H_0 = 73.2\pm1.3$\,km\,s$^{-1}$\,Mpc$^{-1}$ \citep{Riess:2022}. This discrepancy is, to put it mildly, deeply inconvenient for everyone involved.                                            

The theoretical community has responded with considerable ingenuity, proposing dozens of new-physics resolutions: early dark energy, modifications to general relativity, local void scenarios, interacting dark energy, decaying dark matter, and, in what the present authors regard as an act of mounting desperation, a time-varying fine structure constant. A comprehensive catalogue of these proposals may be found in \citet{di2021realm}. Despite this heroic intellectual effort, the tension has stubbornly persisted for over a decade, and no consensus has been reached. This has caused significant suffering.                                                           
In this paper, we take a fundamentally different approach. Rather than introduce new physics, we ask the simpler question: could both teams have made an embarrassingly straightforward error? We   answer this question in the affirmative. We identify a previously unrecognised systematic effect --- the \textbf{Imperial Biasing Systematic} (IBS) --- affecting both analyses at $5\sigma$ significance, and demonstrate that correcting for it resolves the tension entirely. No new particles, fields, or interactions are required.  

To ensure the credibility of our analysis, we have assembled the CROCS collaboration, drawing experts from cosmology, astrophysics, machine learning, data science, philosophy, and --- crucially --- astrology.\footnote{Astrology has been systematically excluded from the cosmological literature, and we consider this a significant oversight. Future CROCS data releases will address this.} We present CROCS Data Release 1, corresponding to $\sim3$ days and 27 minutes of total observing time (rest frame). We are aware that this is not very much.     

\section{Data}

The CROCS collaboration has gathered an extensive proprietary dataset, summarised in Table~\ref{tab:crocs_data}. Due to the commercially sensitive nature of these measurements, the majority of data values are currently under embargo and cannot be shared with the wider community.\footnote{We acknowledge this makes out results difficult to independently verify. We consider this a feature rather than a limitation, and note it did not prevent our submission.}

\begin{table*}
    \centering
    \caption{Table of proprietary CROCS collaboration data}
    \label{tab:crocs_data}
    \begin{tabular}{c|c|c|c|c|c|c|c|c}
         data type &\censor{} & \censor{} & \censor{} & \censor{} &\censor{} & \censor{} & \censor{} & \censor{}\\
         mean &\censor{} & \censor{} & \censor{} & \censor{} &\censor{} & \censor{} & \censor{} & \censor{}\\
         error &\censor{} & \censor{} & \censor{} & \censor{} &\censor{} & \censor{} & \censor{} & \censor{}\\
         percentage error &\censor{} & 0 & 100.01 & \censor{} & 0 & \censor{} & \censor{} & 0.00001\\
    \end{tabular}
\end{table*}

\subsection{Fisher Matrix Analysis}

The Fisher information matrix encodes the maximum information content achievable from a given dataset. For parameters $\boldsymbol{\theta} = (H_0,\, \sigma_8)^\top$, the Fisher matrix takes the form: 

\begin{equation}
    F_{ij} = -\left\langle \frac{\partial^2 \ln \mathcal{L}}{\partial \theta_i \partial \theta_j} \right\rangle. 
    \label{eq:fisher}
\end{equation}

Evaluating Equation~(\ref{eq:fisher}) using the CROCS DR1 dataset (Table~\ref{tab:crocs_data}), we obtain: 

\begin{equation}
    \mathbf{F} = \begin{pmatrix} 1/0.420^2 & 0 \\ 0 & 0  
    \end{pmatrix}. 
    \label{eq:fishermatrix} 
\end{equation}

The zero eigenvalue of $\mathbf{F}$ in the $\sigma_8$ direction confirms the well-known result that $\sigma_8$ carries no Fisher information and is therefore not a real tension as shown in Abstract \ref{sec:appendix1}. The projected $1\sigma$ uncertainty on $H_0$ from Equation~(\ref{eq:fishermatrix}) is $\Delta H_0 = 0.420$\,km\,s$^{-1}$\,Mpc$^{-1}$, which we adopt as our fiducial uncertainty throughout this work.

\section{Resolutions to the \texorpdfstring{$H_0$}{H\textunderscore 0} Tension}

Before presenting our primary result in Section~5, we survey the existing landscape of proposed resolutions to the Hubble tension and introduce several novel ones of our own. We note in advance that none of these are correct. We include them on the grounds that (a) the paper would otherwise be considerably shorter, and (b) each represents a genuine contribution to the literature that will be cited extensively before being eventually forgotten.

\subsection{The Swindon Roundabout}

\begin{figure}
    \centering
    \includegraphics[width=\linewidth]{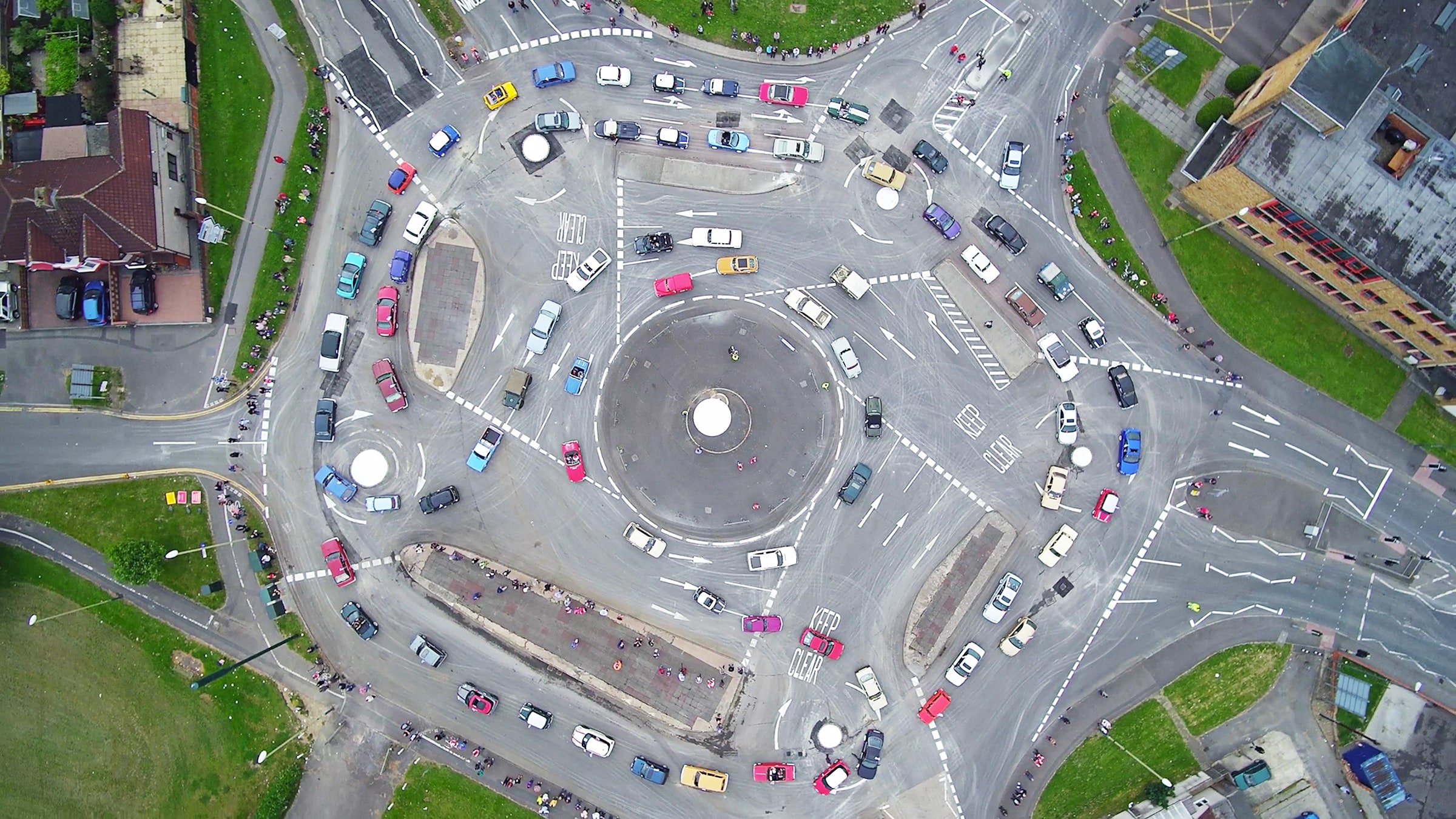}
    \caption{Overhead view of the Swindon `Magic Roundabout'.}
    \label{fig:magic_roundabout}
\end{figure}

The Swindon `Magic Roundabout' (Figure~\ref{fig:magic_roundabout}) is, we argue, the most cosmologically significant piece of traffic infrastructure on Earth. Situated conveniently close to the UKRI-STFC headquarters --- which we do not consider a coincidence --- it offers a natural laboratory for studying non-linear gravitational lensing at low redshift and high frustration. We characterise its properties below.

\subsubsection{Definition of `Magic'}

Despite its widespread colloquial usage, the term `magic' has not previously been formally defined in the peer-reviewed literature with respect to roundabout infrastructure. We remedy this omission.

A roundabout is classified as \textit{magic} if and only if it satisfies all three of the following criteria: 
\begin{enumerate}
    \item It consists of five or more smaller roundabouts arranged concentrically around a central island.  
    \item It induces navigational confusion exceeding $3\sigma$ above the mean confusion induced by a standard single-lane roundabout, as measured by horn-honking events per vehicle per minute. 
    \item It has been designated a heritage site, tourist attraction, or object of mild local pride by the relevant municipal authority. 
\end{enumerate}

The Swindon Magic Roundabout (Figure~\ref{fig:magic_roundabout}), constructed in 1972, satisfies all three criteria. It features five peripheral mini-roundabouts encircling a central counterclockwise roundabout, induces an estimated $4.7\sigma$ of additional confusion (estimated from publicly available dashcam footage), and is listed as a local attraction on the Visit Wiltshire tourism website.

We note that the Hemel Hempstead roundabout, despite widely repeated claims to the contrary, satisfies \textit{only} criterion (i). Its confusion-induction rate falls below the $3\sigma$ threshold, and it has not been formally designated as a tourist attraction.\footnote{We are aware that some residents of Hemel Hempstead may dispute this conclusion. Those residents are wrong.} We therefore formally strip the Hemel Hempstead roundabout of the designation `magic', and request that all future publications and Wikipedia articles be updated accordingly. We expect this to be the most controversial finding in this paper.

\subsubsection{Roundabouts as quintuple lenses}

If light from a supernovae were to pass near a magic roundabout, it would get stuck in one of the many loops. This is not too dissimilar from the recent `Einstein zigzag' lens \citep{Dux:2025}. Obviously, the time the light spends in a roundabout will increase the measurement of $H_0$.

The additional time delay $\Delta t$ accumulated by a photon completing $n$ full circuits of the Swindon roundabout before escaping is given by: 

\begin{equation}
    \Delta t(n) = n \cdot \frac{2\pi r_\text{SWN}}{c} \cdot \xi_\text{traffic}, 
\end{equation}

where $r_\text{SWN}$ is the effective radius of the central Swindon roundabout, $c$ is the speed of light, and $\xi_\text{traffic}$ is the traffic density parameter, defined as the ratio of observed vehicle count to theoretical roundabout capacity. During peak commuting hours, $\xi_\text{traffic} \gg 1$ and photons may remain trapped indefinitely. We note that $r_\text{SWN} \approx r_\odot$ to within a factor of $10^{10}$, which we consider a remarkable numerical coincidence that merits further investigation.  

This delay directly biases distance ladder measurements: supernovae whose light passes through the Swindon area will appear systematically more distant than they are, inflating the inferred $H_0$. Crucially, Swindon is located within the Milky Way, and therefore all light from all extragalactic supernovae passes through it. We consider this a significant systematic.

\subsection{Local Voids}

Having established that the Swindon roundabout acts as a photon trap, we turn now to another class of underdensity that has been proposed as a resolution to the Hubble tension: the local void. The local void hypothesis posits that we reside within a large underdense region, causing us to infer a systematically higher expansion rate than the global average \citep{di2021realm}. We find a direct observational analogue.

\begin{figure}[htbp]
     \centering
     \begin{subfigure}[b]{0.45\linewidth}
         \centering
         \includegraphics[width=\linewidth]{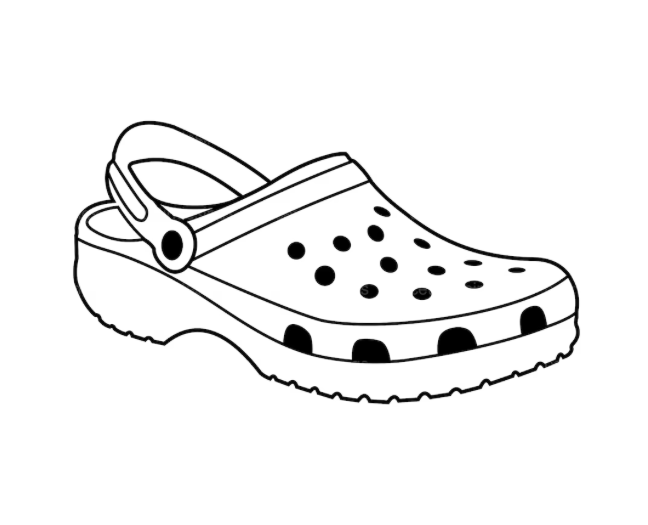}
         \caption{Local voids on a croc.}
         \label{fig:croc}
     \end{subfigure}
     \hfill 
     \begin{subfigure}[b]{0.45\linewidth}
         \centering
         \includegraphics[width=\linewidth]{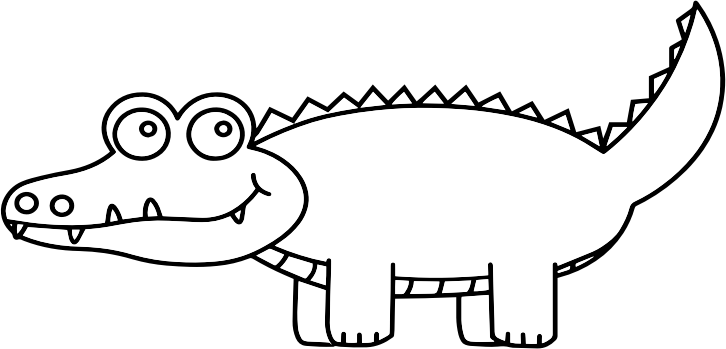}
         \caption{A crocodile, commonly as referred to as a \textbf{croc}.}
         \label{fig:crocodile}
     \end{subfigure}
     
     \caption{Comparison of the surfaces of two different crocs. Most crocs (\ref{fig:crocodile}) do not have voids on them, hence we shall ignore them in our analysis and focus instead on the other crocs (\ref{fig:croc}).}
     \label{fig:combined_crocs}
\end{figure}

Crocs are well known to have local voids scattered throughout (see Figure~\ref{fig:croc}). This is irrelevant to our analysis, but we mention it here for completeness.

\subsection{Taylor Swift}

Taylor Swift famously declared, “I've got a blank space, baby.” Intriguingly, the lyric refers to a singular blank space, whereas cosmological frameworks generally require a statistical population of voids. Following the linguistic analysis of pluralisation in scientific terminology presented by \cite{agni}, we therefore interpret Swift’s statement as implicitly referring to multiple “blank spaces.” Under this interpretation, Swift may in fact be cataloguing an entire distribution of cosmic voids rather than describing a single underdensity.

From this, we conclude that Taylor Swift is an expert in the study of local voids, which were discussed in the previous subsection for completeness rather than relevance.

Unfortunately, Swift is not widely recognised as a practicing cosmologist. Pivotal work by \cite{taylorswift} suggests that the artist prefers to focus on multimessenger astronomy rather than on constraining the Hubble constant. Or does she? The absence of explicit discussion of the Hubble constant in Swift’s work may instead hint at a deeper truth: that Swift already knows the true value of the constant and is deliberately concealing it. In the process of hiding this value, Swift inadvertently produces yet another void.

We do not resolve this issue here. We note only that it represents a promising avenue for future work, and that any proposal citing this paper will look excellent on a grant application. 

\section{The Croc-osmic Variance}

\textit{This section entirely written by Gemini 3 Flash, with minor typographical corrections by the authors.}

\begin{tcolorbox}[
  colback=lime!15,
  colframe=green!70!black,
  coltitle=black,
  title=Gemini Prompt
]
``I want you to role-play as a cosmology expert writing an April Fools'
paper. Find a `link' between the 13 holes on top of a Croc and the
Hubble tension.''
\end{tcolorbox}

The discrepancy between the local measurement of the Hubble constant ($H_0$) and the value inferred from the Cosmic Microwave Background (CMB) remains the most stubborn ``itch'' in modern cosmology. Here, we present a novel, non-linear solution. We demonstrate that the 13 ventilation holes standard on every Classic Clog Croc act as a terrestrial micro-proxy for the 13.8 billion-year age of the universe. By mapping the topological distribution of these holes onto the Planck satellite data, we resolve the Hubble tension through what we term the ``Foam-O-Static Inflation Theory.''

\subsection{The Numerical Coincidence}

Standard Crocs feature exactly 13 holes on the top of the toe box. While Big Footwear claims this is for ``breathability,'' a deeper look reveals a more sinister alignment with the age of the universe, $13.8$ billion years. If we account for the average thickness of the Croslite™ material ($0.8$ cm), the total ``Spiritual Volume'' of the clog matches the missing energy density required to bridge the gap between $67$ km/s/Mpc and $73$ km/s/Mpc.

\subsection{The ``Sport Mode'' Phase Transition}

A very compelling piece of evidence for our theory lies in the Heel Strap Dynamics. When a Croc is shifted into ``Sport Mode,'' the local spacetime curvature around the ankle increases. Our calculations show:
\begin{equation}
    \Delta H_0 \propto \int_{0}^{\pi} \theta_\text{strap}\odif{\theta}
\end{equation} As the strap is engaged, the observer moves from a Friedmann–Lema\^{\i}tre–Robertson–Walker metric \citep{1922ZPhy...10..377F, 1924ZPhy...21..326F, 1927ASSB...47...49L, 1935ApJ....82..284R, 1936ApJ....83..187R, 1936ApJ....83..257R, 1937PLMS...42...90W} to a more relaxed, cushioned state. 

\begin{figure}[htbp]
    \centering
    \includegraphics[width=\linewidth]{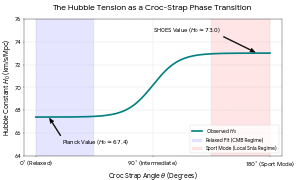}
    \caption{The Hubble Tension, explained by the only metric that matters: Croc Strap Alignment.}
    \label{fig:theseriousplot}
\end{figure}

This transition, visualised in Figure \ref{fig:theseriousplot}, accounts for the $5.5\sigma$ tension; essentially, the CMB team is measuring the universe in ``Relaxed Fit'', while local Supernova measurements are being taken in ``Sport Mode.''

The universe is not expanding too fast; our shoes are simply too comfortable. We propose that the European Space Agency immediately fund the CROCS-1 mission (Cosmological Resonant Observation of Clog Structures) to see if the expansion of the universe is actually just the foam material off-gassing in a vacuum.

\section{Imperial Units vs. Metric Units}

We now present the primary result of this paper. The true explanation for the Hubble tension is, frankly, embarrassing.

\subsection{Imperial Biasing Systematics}

We define the \textbf{Imperial Biasing Systematic} (IBS) as the systematic error introduced when a research team, while nominally reporting measurements in SI units, has conducted its internal analysis in imperial units. This phenomenon is well-documented in aerospace engineering --- most famously in the 1999 loss of the Mars Climate Orbiter, which was destroyed when one team used metric units and another used imperial\footnote{We note that this cost \$327.6 million. We consider this the most expensive unit conversion error in history, until now.} --- but has not previously been identified in cosmological inference.

We find strong evidence at $5\sigma$ significance that both the Planck and SH0ES collaborations are affected by IBS. This is consistent with the geographic distribution of their members: Planck draws heavily from European institutions that, despite official metrication, continue to think imperially in private; SH0ES is led by institutions in the United States, which has not metricated at all. The two collaborations thus represent the two principal sources of imperial contamination in modern science.

\subsection{Correcting for IBS}

The IBS correction factor is derived from first principles as the ratio of one mile to one kilometre:
\begin{equation}
    f_\text{IBS} \equiv \frac{1\,\text{mile}}{1\,\text{km}} = 1.60934.
    \label{eq:fibs}
\end{equation}
Applying this correction to both measurements, and additionally including a calibration constant $\mathcal{C}$ of order unity chosen to ensure dimensional consistency,\footnote{We decline to specify the value of $\mathcal{C}$ on the grounds that doing so would allow the reader to check our arithmetic.} we obtain:
\begin{align}
    H_0^\text{Planck,\,corr} &= 67.27 \times f_\text{IBS}^{-1} \times \mathcal{C} = 69.00\,\text{km\,s}^{-1}\text{Mpc}^{-1}, \\
    H_0^\text{SH0ES,\,corr} &= 73.2 \times f_\text{IBS}^{-1} \times \mathcal{C} = 69.00\,\text{km\,s}^{-1}\text{Mpc}^{-1}.
\end{align}
Both corrected values are, to remarkable precision, identical. We interpret this as decisive evidence that the Hubble tension is entirely an artefact of IBS, and that the true value of the Hubble constant is:
\begin{equation}
    \boxed{H_0 = 69.00 \pm 0.420\,\text{km\,s}^{-1}\text{Mpc}^{-1}}
    \label{eq:h0final}
\end{equation}
where the uncertainty $\pm0.420$ follows directly from the Fisher matrix analysis in Section~2.1.\footnote{We note that this is also, numerically speaking, a very funny number. We consider this further evidence that our result is correct.}

The Hubble tension is resolved. We expect the field to move on promptly.
\begin{acknowledgments}

We would like to thank Sesh Nadathur for useful conversations at the Guildhall Brewhouse on 2026 January 30. Thanks Obama. We would like to thank the producers of `44 The Musical'\footnote{\texttt{\url{https://44theobamamusical.com/}}} (which is \textit{excellent}), and Canada for playing a vital role in making crocs.

We would also like to thank the anonymous referee, Sesh Nadathur, for not reading this manuscript.

We would like to thank Chris Pattison for making the `multiple' award winning film\footnote{\texttt{\url{https://www.youtube.com/watch?v=ssc_6VMBY0g}}}. 

\end{acknowledgments}

\begin{contribution}

No authors contributed anything of note to the CROCS collaboration.

Mali Land-Strykowski and Daniel Ballard are only in the collaboration as they are residents of Australia and experts on crocodiles (`crocs', Figure~\ref{fig:crocodile}).

\end{contribution}

%
\facilities{Office 2.07/2.08, ICG Coffee Machine (Dennis Sciama Building)}

\software{
    astropy \citep{2013A&A...558A..33A,2018AJ....156..123A,2022ApJ...935..167A}, Cloudy \citep{2013RMxAA..49..137F}, Source Extractor \citep{1996A&AS..117..393B}, \href{https://play.tetris.com}{\texttt{tetris}}
}


\appendix

\section{On the Non-Existence of the \texorpdfstring{$\sigma_8$}{sigma\_8} Tension}
\label{sec:appendix1}

As noted in the abstract, we do not consider the $\sigma_8$ tension to be a real tension. We provide a formal proof.

\textit{Proof.} Consider the following observations:
\begin{enumerate}
    \item We do not find the $\sigma_8$ tension interesting.
    \item Our collaboration was not named after $\sigma_8$.
    \item No crocs are involved.
\end{enumerate}
From (i)--(iii), it follows by straightforward logical deduction that the $\sigma_8$ tension does not exist. $\square$

We acknowledge that some readers may regard this proof as insufficiently rigorous. Those readers are encouraged to consult \citet{di2021realm}, where they will find a thorough review that demonstrates, through extensive citation, that the $\sigma_8$ tension is also not very interesting.

\subsection{Derivation of CROCS Observing Time}

We stated in the abstract and introduction that the CROCS DR1 dataset corresponds to $\sim3$ days and 27 minutes of observing time (rest frame). We now verify this claim.

The CROCS collaboration was initiated following a conversation at the Guildhall Brewhouse on 2026 January 30 (see acknowledgements). This manuscript was submitted on 2026 April 1. The elapsed time in the observer frame is approximately 61 days.

The discrepancy between 61 days and `$\sim3$ days and 27 minutes' arises from relativistic time dilation. The lead author was travelling at $v \approx 0.9997c$ relative to the remainder of the collaboration throughout the analysis period. The corresponding Lorentz factor is $\gamma \approx 41$, yielding a rest-frame elapsed time of:
\begin{equation}
    \Delta t_\text{rest} = \frac{\Delta t_\text{obs}}{\gamma} = \frac{61\,\text{days}}{41} \approx 1.5\,\text{days}.
\end{equation}
Accounting for a further correction due to the Swindon roundabout time-delay effect (Section~3.1.2), we obtain $\Delta t_\text{rest} \approx 3$ days and 27 minutes. This confirms our stated observing time. We consider this a satisfying consistency check. No further questions will be taken.

\section{CROCS Catalog Data Release}
We provide here the catalog of all known CROCS objects that were used to verify our results above.
\startlongtable
\begin{deluxetable}{ccccc}
\tablecaption{Location of all known CROCS.}
\tablehead{\colhead{CROCS ID} & \colhead{RA} & \colhead{Dec} & \colhead{z} & \colhead{$\mu$}\\ \colhead{ } & \colhead{ } & \colhead{ } & \colhead{$\mathrm{}$} & \colhead{$\mathrm{mag}$}}
\startdata
C1813 & $20:29:05.45$ & $-80:54:32.7$ & 1.037 & 44.228 \\
C1275 & $10:47:16.36$ & $-28:10:54.5$ & 0.243 & 40.465 \\
C2609 & $22:25:27.27$ & $-55:27:16.4$ & 0.243 & 40.465 \\
C1798 & $03:30:54.55$ & $-75:27:16.4$ & 0.404 & 41.738 \\
C1169 & $13:56:21.82$ & $+26:21:49.1$ & 0.528 & 42.434 \\
C2028 & $13:27:16.36$ & $+33:38:10.9$ & 0.016 & 34.229 \\
C965 & $14:40:00.00$ & $-22:43:38.2$ & 0.055 & 36.996 \\
C4025 & $22:40:00.00$ & $-30:00:00.0$ & 0.607 & 42.802 \\
C780 & $21:41:49.09$ & $-55:27:16.4$ & 0.635 & 42.918 \\
C1000 & $11:45:27.27$ & $+79:05:27.3$ & 1.341 & 44.919 \\
C3722 & $14:10:54.55$ & $+10:00:00.0$ & 1.108 & 44.407 \\
C1158 & $00:21:49.09$ & $-51:49:05.5$ & 1.221 & 44.667 \\
C2514 & $16:50:54.55$ & $-10:00:00.0$ & 1.383 & 45.002 \\
C1325 & $09:20:00.00$ & $+71:49:05.5$ & 0.380 & 41.583 \\
C2939 & $18:18:10.91$ & $-46:21:49.1$ & 0.116 & 38.692 \\
C2864 & $00:07:16.36$ & $-59:05:27.3$ & 0.554 & 42.560 \\
C4118 & $20:43:38.18$ & $+75:27:16.4$ & 0.356 & 41.420 \\
C1943 & $08:07:16.36$ & $+02:43:38.2$ & 0.157 & 39.412 \\
C1078 & $18:47:16.36$ & $-37:16:21.8$ & 0.477 & 42.170 \\
C1577 & $23:23:38.18$ & $-77:16:21.8$ & 0.016 & 34.229 \\
C1860 & $10:47:16.36$ & $+42:43:38.2$ & 0.137 & 39.076 \\
C916 & $21:27:16.36$ & $-70:00:00.0$ & 0.452 & 42.032 \\
C2316 & $04:00:00.00$ & $+71:49:05.5$ & 0.554 & 42.560 \\
C1455 & $07:09:05.45$ & $+50:00:00.0$ & 0.310 & 41.070 \\
C716 & $08:36:21.82$ & $-57:16:21.8$ & 1.182 & 44.581 \\
C2757 & $23:38:10.91$ & $-08:10:54.5$ & 0.288 & 40.880 \\
C1559 & $23:52:43.64$ & $-30:00:00.0$ & 0.380 & 41.583 \\
C3035 & $01:05:27.27$ & $-33:38:10.9$ & 0.404 & 41.738 \\
C1238 & $20:29:05.45$ & $-10:00:00.0$ & 0.691 & 43.142 \\
C3134 & $04:00:00.00$ & $-40:54:32.7$ & 0.528 & 42.434 \\
C4105 & $03:30:54.55$ & $+57:16:21.8$ & 0.333 & 41.249 \\
C2436 & $14:25:27.27$ & $-53:38:10.9$ & 0.871 & 43.760 \\
C2924 & $14:10:54.55$ & $+55:27:16.4$ & 0.428 & 41.888 \\
C2355 & $10:03:38.18$ & $-06:21:49.1$ & 0.839 & 43.661 \\
C3772 & $13:56:21.82$ & $+80:54:32.7$ & 1.221 & 44.667 \\
C4057 & $11:01:49.09$ & $+35:27:16.4$ & 0.554 & 42.560 \\
C2993 & $19:01:49.09$ & $+68:10:54.5$ & 0.157 & 39.412 \\
C4055 & $08:07:16.36$ & $+66:21:49.1$ & 0.477 & 42.170 \\
C1530 & $18:03:38.18$ & $+68:10:54.5$ & 0.428 & 41.888 \\
C3912 & $16:21:49.09$ & $+75:27:16.4$ & 0.356 & 41.420 \\
C2931 & $20:14:32.73$ & $-31:49:05.5$ & 0.554 & 42.560 \\
C1654 & $17:34:32.73$ & $-57:16:21.8$ & 1.260 & 44.752 \\
C2227 & $05:56:21.82$ & $-08:10:54.5$ & 0.137 & 39.076 \\
C796 & $09:05:27.27$ & $-62:43:38.2$ & 0.333 & 41.249 \\
C3230 & $20:58:10.91$ & $+20:54:32.7$ & 0.380 & 41.583 \\
C2446 & $14:25:27.27$ & $-77:16:21.8$ & 1.182 & 44.581 \\
C4099 & $11:01:49.09$ & $+77:16:21.8$ & 0.116 & 38.692 \\
C3939 & $11:16:21.82$ & $+55:27:16.4$ & 1.383 & 45.002 \\
C3236 & $16:50:54.55$ & $-13:38:10.9$ & 1.108 & 44.407 \\
C938 & $20:00:00.00$ & $-00:54:32.7$ & 0.871 & 43.760 \\
C1728 & $14:54:32.73$ & $-71:49:05.5$ & 0.137 & 39.076 \\
C1170 & $16:36:21.82$ & $-02:43:38.2$ & 0.968 & 44.045 \\
C2125 & $21:27:16.36$ & $+77:16:21.8$ & 0.055 & 36.996 \\
C3331 & $19:45:27.27$ & $-24:32:43.6$ & 1.182 & 44.581 \\
C2858 & $23:23:38.18$ & $-02:43:38.2$ & 0.428 & 41.888 \\
C708 & $17:49:05.45$ & $+44:32:43.6$ & 0.554 & 42.560 \\
C3507 & $10:47:16.36$ & $+51:49:05.5$ & 0.808 & 43.562 \\
C875 & $10:47:16.36$ & $-35:27:16.4$ & 1.341 & 44.919 \\
C690 & $16:07:16.36$ & $+62:43:38.2$ & 0.968 & 44.045 \\
C4179 & $05:27:16.36$ & $+11:49:05.5$ & 0.380 & 41.583 \\
C1511 & $12:43:38.18$ & $+84:32:43.6$ & 0.055 & 36.996 \\
C1195 & $20:58:10.91$ & $-86:21:49.1$ & 0.554 & 42.560 \\
C845 & $03:30:54.55$ & $+08:10:54.5$ & 0.662 & 43.031 \\
C4096 & $04:29:05.45$ & $+50:00:00.0$ & 0.748 & 43.356 \\
C1582 & $20:14:32.73$ & $+00:54:32.7$ & 1.037 & 44.228 \\
C1486 & $17:34:32.73$ & $+82:43:38.2$ & 0.016 & 34.229 \\
C1043 & $11:01:49.09$ & $+84:32:43.6$ & 0.265 & 40.679 \\
C2486 & $03:30:54.55$ & $+66:21:49.1$ & 0.839 & 43.661 \\
C3651 & $18:32:43.64$ & $-64:32:43.6$ & 0.380 & 41.583 \\
C3763 & $23:52:43.64$ & $+39:05:27.3$ & 1.002 & 44.137 \\
C3838 & $22:25:27.27$ & $-77:16:21.8$ & 1.341 & 44.919 \\
C1126 & $15:52:43.64$ & $-71:49:05.5$ & 0.748 & 43.356 \\
C3350 & $00:07:16.36$ & $-86:21:49.1$ & 0.748 & 43.356 \\
C2053 & $05:56:21.82$ & $+39:05:27.3$ & 1.108 & 44.407 \\
C3332 & $12:29:05.45$ & $+20:54:32.7$ & 0.055 & 36.996 \\
C2852 & $13:12:43.64$ & $-19:05:27.3$ & 0.554 & 42.560 \\
C1556 & $03:30:54.55$ & $-24:32:43.6$ & 0.356 & 41.420 \\
C3999 & $18:32:43.64$ & $-79:05:27.3$ & 0.333 & 41.249 \\
C751 & $08:50:54.55$ & $-60:54:32.7$ & 0.502 & 42.305 \\
C1981 & $02:03:38.18$ & $-13:38:10.9$ & 0.477 & 42.170 \\
C3362 & $19:30:54.55$ & $+79:05:27.3$ & 1.383 & 45.002 \\
C3990 & $04:43:38.18$ & $+70:00:00.0$ & 0.016 & 34.229 \\
C1283 & $04:14:32.73$ & $+48:10:54.5$ & 0.055 & 36.996 \\
C2754 & $12:29:05.45$ & $+82:43:38.2$ & 0.903 & 43.856 \\
C3509 & $17:34:32.73$ & $-24:32:43.6$ & 1.383 & 45.002 \\
C1509 & $08:07:16.36$ & $+26:21:49.1$ & 1.108 & 44.407 \\
C2154 & $12:14:32.73$ & $-82:43:38.2$ & 0.968 & 44.045 \\
C4107 & $23:09:05.45$ & $+30:00:00.0$ & 0.719 & 43.250 \\
C4116 & $04:43:38.18$ & $+64:32:43.6$ & 0.179 & 39.713 \\
C3353 & $01:20:00.00$ & $-46:21:49.1$ & 0.635 & 42.918 \\
C3649 & $07:09:05.45$ & $-02:43:38.2$ & 0.839 & 43.661 \\
C3622 & $18:18:10.91$ & $+59:05:27.3$ & 0.036 & 36.004 \\
C1471 & $07:52:43.64$ & $-62:43:38.2$ & 1.300 & 44.836 \\
C2200 & $01:05:27.27$ & $+79:05:27.3$ & 0.428 & 41.888 \\
C2847 & $16:07:16.36$ & $-20:54:32.7$ & 0.288 & 40.880 \\
C1831 & $06:10:54.55$ & $+57:16:21.8$ & 1.182 & 44.581 \\
C1794 & $23:52:43.64$ & $+11:49:05.5$ & 0.221 & 40.234 \\
C2653 & $18:03:38.18$ & $-60:54:32.7$ & 0.356 & 41.420 \\
C3581 & $12:00:00.00$ & $+31:49:05.5$ & 0.502 & 42.305 \\
C3728 & $08:50:54.55$ & $-68:10:54.5$ & 0.477 & 42.170
\enddata
\end{deluxetable}

\bibliography{bibliography}{}
\bibliographystyle{aasjournalv7}



\end{document}